\documentclass{aastex}
\usepackage{emulateapj5}

\newcommand{\etal}{{et al.}}

\slugcomment{accepted 2001-09-20}

\shorttitle{New wide pair of cool white dwarfs}
\shortauthors{Scholz et al.}

\begin{document}


\title{A new wide pair of cool white dwarfs in the Solar neighbourhood}


\author{R.-D. Scholz, G. P. Szokoly and M. Andersen }
\affil{Astrophysikalisches Institut Potsdam,
    An der Sternwarte 16, D-14482 Potsdam, Germany}

\author{R. Ibata}
\affil{Observatoire de Strasbourg, 11, rue de l'Universite,
       F-67000 Strasbourg, France}

\author{M. J. Irwin}
\affil{Institute of Astronomy, Madingley Road, Cambridge CB3 0HA, UK}


\begin{abstract}
We report the discovery of a wide pair (93~arcsec angular separation) of 
extremely cool ($T_{eff}<4000$~K) white dwarfs with a very large common 
proper motion ($\sim1.9$~arcsec/yr). The objects were discovered in a high 
proper motion survey in the poorly investigated southern sky region with 
$\delta<-60^{\circ}$ using SuperCOSMOS Sky Survey (SSS) data.  Both objects, 
SSSPM~J2231-7514 and SSSPM~J2231-7515, show featureless optical spectra. 
Fits of black-body models to the spectra yield effective temperatures of 
3810~K and 3600~K, respectively for the bright ($V=16.60$) and faint 
($V=16.87$) component. Both degenerates are much brighter than other recent 
discoveries of cool white dwarfs with comparable effective temperatures
and/or $B_J-R$ colours. 
Therefore, they should be relatively nearby objects.  
The comparison with other cool white dwarfs and a photometric
distance determination yield distance estimates between 9~pc
and 14~pc. The latter seems to be more realistic, since
the good agreement of the proper motion of both components
within the errors of about 8~mas/yr and the angular separation between the 
two stars support a distance of about 15~pc with relatively small masses of
the components. 
With smaller distance we should be able to measure a differential proper
motion due to orbital motion if the orbital plane is not strongly inclined
and the present orbital velocity vector is not close to the line of sight.
The space velocity based on that distance and assumptions on radial velocity
makes the new pair of extremely cool white dwarfs some of the probably oldest
members of the Galactic disk population, although the possibility that
these objects are part of a Galactic halo dark matter component can also
not yet be ruled out.
\end{abstract}




\section{Introduction}

Cool white dwarfs are interesting objects for a number of reasons such as 
determining the age of the Galactic disk \citep{leg98} and as possible 
components of the Galactic dark matter halo \citep{iba00,opp01b}.
A detailed photometric and spectroscopic analysis of all known cool white 
dwarfs (4000~K$<T_{eff}<12000$~K) with trigonometric parallax measurements was 
summarised by \citet{ber01}. Additionly there are a number of recent 
discoveries of cool white dwarfs with effective temperatures below
4000~K \citep{ham97,iba00,har01,opp01b}, one of which, WD~0346$+$246, has
a trigonometric parallax measurement \citep{ham99}.

As shown by \citet{opp01b}, the temperature of cool white dwarfs
is well correlated with the $B_J-R$ colour of photographic plates
taken with the United Kingdom Schmidt Telescope (UKST).  \citet{opp01b} have 
also presented a photographic colour-colour diagram $(B_J-R)$ vs. $(R-I)$
of cool white dwarfs together with curves of synthetic colours expected
for different chemical compositions. These observations can now be
used for confirmation of recent theoretical models on the formation of 
molecular hydrogen in cool white dwarf atmospheres leading to collisionally-
induced opacity at wavelengths longer than 6000{\AA} \citep{han98}. 

\section{Discovery of the common high proper motion pair}

Several high proper motion surveys have been carried out recently
with the aim of completing the proper motion catalogues of Luyten, e.g. 
\citet{luy79}, at fainter magnitudes and finding new low-luminosity 
objects \citep{mon00,sch00,rui01,opp01b}.
Most of these efforts are concentrated on the southern sky with 
$\delta < -33^{\circ}$, where the main source for Luyten's catalogues is
the Bruce Proper Motion Survey, most of whose plates reach only to about 
$m_{pg} = 15.5$. In the northern sky, Luyten's catalogues are based on
observations with the Palomar Schmidt telescope.

Recently, the entire southern sky has been scanned in two passbands ($B_J$ 
and $R$), with additional scans of plates at different epochs and in other 
passbands (including $I$) over large sky areas \citep{ham01a,ham01b,ham01c} -
(hereafter SSS).  SSS data were the 
observational basis of the recent survey for cool white dwarfs by 
\citet{opp01b}, covering about 10\% of the sky around the South Galactic Cap.
We have made use of the same observational material but our search was 
carried out in a region close to the south celestial pole, at
declinations $\delta < -60^{\circ}$, where in most fields three different
epoch measurements on UKST $B_J$ and $R$ plates and ESO $R$ plates were
available.

The basic reduction consisted of a re-matching of previously (i.e. within a
standard identification radius of 3~arcsec) unmatched objects and the 
subsequent determination of proper motions using linear regression over
the SSS measured coordinates at three epochs.
The search area is difficult to estimate as well as the completeness
of the survey. This is due to the fact that in many fields the epochs
of two of the plates were very close to each other. On the other hand we
made a rather restrictive preselection of undisturbed star-like
objects based on their image parameters, in order to exclude spurious
matching leading to false proper motions. Therefore, we do not attempt to
draw conclusions on the statistical properties of the survey, but just present
the discovery of the two objects with the largest proper motion obtained
so far in the SSS data.

The coordinates, photographic photometry and the
proper motions of the two high proper motion stars are given in 
Table~\ref{sssdata}. The $I$ band magnitudes (as well as the corresponding 
extra-epoch positions used in the proper motion solution) were kindly provided 
to us by Nigel Hambly. A further improvement of the proper motion solution
was achieved from including additional positions measured on the recent epoch 
acquisition images. These positions were obtained in the system of the SSS
catalogues by using all stars with small proper motions in a field of 
$15 \times 15$~arcmin$^2$ as reference stars (see note to Table~\ref{sssdata}).
There is no significant difference between the proper motions of the two
components. Since the discovery of the wide pair of cool white dwarfs 
with very large common proper motion
is based on SSS data, we have given them the names SSSPM~J2231-7514 and
SSSPM~J2231-7515.

SSSPM~J2231-7515 (but not its companion since it was brighter than the survey
limit) was also discovered in another on-going program using APM measures of 
the same UKST sky survey plates \citep{iba00}, which also aims to find cool 
high proper motion white dwarfs.

\section{Observations}

\subsection{DFOSC observations}

Observations of the new common proper motion pair
were carried out using the {\em Danish Faint Object
Spectrograph} ({\em DFOSC}) on the Danish 1.54m Telescope in La
Silla. Data were taken during the nights starting June 19-20, 2001 (local
time) in relatively good almost photometric conditions.

\begin{figure}
\plotone{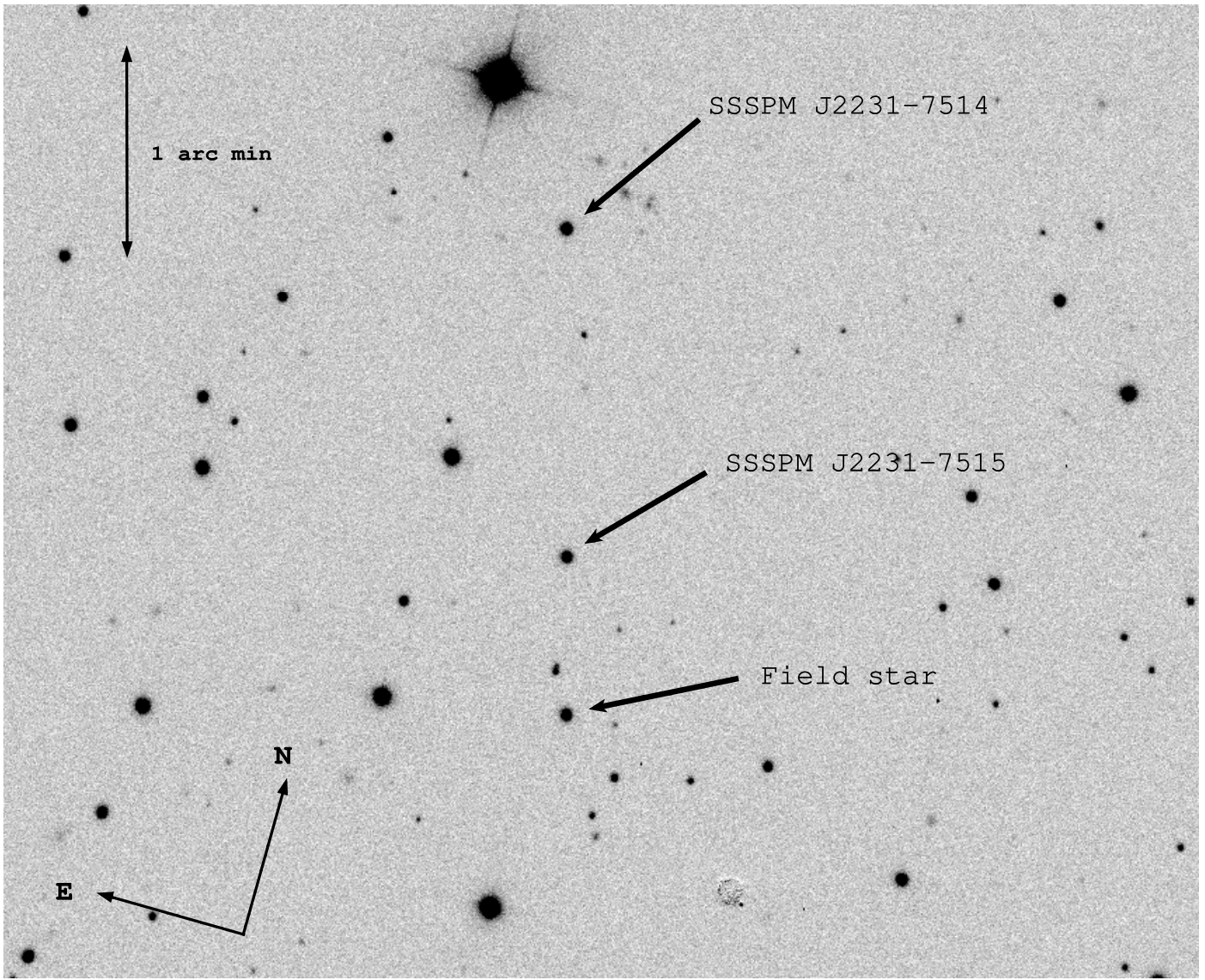}
\caption{The $V$ band acquisition image taken at epoch 2001.47, where
the components of the high proper motion pair are marked together with
a field K-star which was exactly on
the slit and could be used for an additional check of the flux
calibration of the spectra.\label{ssspm_ima}}
\end{figure}

Seeing varied slightly during these nights between 1.0 and 1.2~arcsec.
Only short V-band (Bessel) images were taken that are relevant to our
objects as acquisition images (Figure~\ref{ssspm_ima}). 
Spectroscopy was done using a 2.0~arcsec
wide long-slit (corresponding to 5.1 pixels in imaging mode),
positioning both objects on the slit at the same time. Two
grisms were used, number-5 (5200-10200{\AA}, 7000{\AA} blaze, 3.3{\AA}/pixel
resolution -- one 1800s spectroscopic observation) and number-7
(3800-6800{\AA}, 5250{\AA} blaze, 1.65{\AA}/pixel resolution -- 
a 1800s and a 1300s spectroscopic observation).
Two spectrophotometric standards were taken using each grism, 
LTT 7379 and LTT 9239 \citep{ham92,ham94}, as well as a large number of
zero, flat (both for imaging and spectroscopy for all grisms used) and
arc-lamp (for both grisms) calibration frames. For broad-band photometric
calibrations, Landolt standard fields \citep{lan92} were observed repeatedly.

\subsection{EFOSC observations}

After completing the reduction of the DFOSC observations, we realised 
that a spectrum of one of the objects, SSSPM~J2231-7515, which was discovered
independently, had already been observed half a year earlier.
The spectrum of this star was observed with the EFOSC
spectrograph on the ESO~3.6m telescope during the night of 2 December
2000. A slit width of 1.5~arcsec was used with grism number~1 (3185-10940{\AA},
4500{\AA} blaze, 6.30{\AA}/pixel resolution) for three exposures of 300s each.

\section{Data reduction}

All of the data was reduced in the standard manner using IRAF procedures.
Consequently in the following sections we only describe important results
or deviations from normal.

\subsection{DFOSC data}

For the imaging data all images were bias-corrected, trimmed and then 
flat-fielded using twilight flats.  After measuring photometric zeropoints 
in 6 standard fields \citep{lan92}, we adopted a zeropoint of 24.66 (for 1s 
exposure time and airmass of 1.45 -- the airmass of our acquisition frames) 
in the Bessel V-band. Examination of the measured zeropoints shows that 
conditions were very close to photometric (with a hint of very weak cirrus in
some cases). Using this zeropoint, we measured the (Vega) magnitudes of the
two objects to be 16.60 and 16.87 in V (Bessel). We also measured the magnitude
of the extra object, accidentally landing on our 2 arcsec wide slit (see below)
to be 16.86 (see Figure~\ref{ssspm_ima}).  
Instrumental magnitude errors are small (1-2\%), the 
main source of error is the determination of the zeropoint. We therefore 
estimate the overall accuracy of these magnitudes to be better than 5\%.

Spectroscopic frames (science and standard fields and calibration frames)
were also bias and zero subtracted, trimmed and flatfielded (using different
flat-field frames for the two grisms). The only complication was the 
removal of focal plane geometric distortions to improve the sky subtraction.
This was done by tracing lines in the wavelength calibration frames. A
smooth distortion map was fitted to the results and was applied to all frames.

All objects on the slit were traced along the dispersion axis. Sky subtraction
was done through fitting in a 35~arcsec wide band, centered on the object,
excluding the central 16~arcsec region. A relatively wide aperture was
defined for all objects, which includes all the flux (but degrades the
S/N slightly). Object spectra were `optimally' extracted within this aperture,
with both cosmic-ray removal (based on photon statistics) and a weighted
sum based on estimated signal-to-noise ratio.

Wavelength calibration was done using He-Ne arc exposures. For the 
number-7 grism (bluer, higher resolution) the procedure worked quite
well (0.06{\AA} RMS, 0.15{\AA} maximum deviation) in the wavelength range
3889-6717{\AA}. For the number 5 grism (red, lower resolution), the accuracy
is worse (as expected), around 0.5{\AA} RMS (peak deviation is 1{\AA}), in
the range of 5852-8783{\AA}. In addition, the exact position of the
object on the slit can also introduce an additional, systematic shift of
the spectra. This later effect is estimated to be 1.5{\AA} for the lower
resolution and 0.8{\AA} for the higher resolution grism. These estimates
were also verified by checking emission lines in the sky background.

It is important to note that beyond 7000{\AA} the CCD suffers from serious
fringing, which we could not remove.

Flux calibration was attempted by observing two standard stars in the same
two grism configurations.  However, during the reduction we encountered
several problems with the red end calibration possibly related to the fringing.
The serendipitously observed 3rd object in the slit turned out to be a
K0-K1 star.  Using a library spectrum \citep{jac84}, we corrected the
continuum slope of this object and used this smooth correction to reduce the
calibration problems for the two white dwarfs.  The quality of this correction
was checked by comparing overlapping regions in wavelength in different 
observations.  For the K-star, differences were less than 1\%. For the fainter
high proper motion object (which is closer to the K-star), differences were 
1-2\%, showing no colour dependence. For the brighter high proper motion 
object, differences were 5\%. These residual differences were assumed to be
`grey' effects (i.e. not depending on wavelength). 

Finally we calculated synthetic Bessel-V magnitudes for the objects
from the spectra. In the case of the brighter object, the measured broad-band
magnitude was reproduced better than 1\%, while for the fainter one a 9\%
difference was measured. We applied a (colour independent) correction to
match the spectrum to the measured broadband magnitudes.

\begin{figure}
\plotone{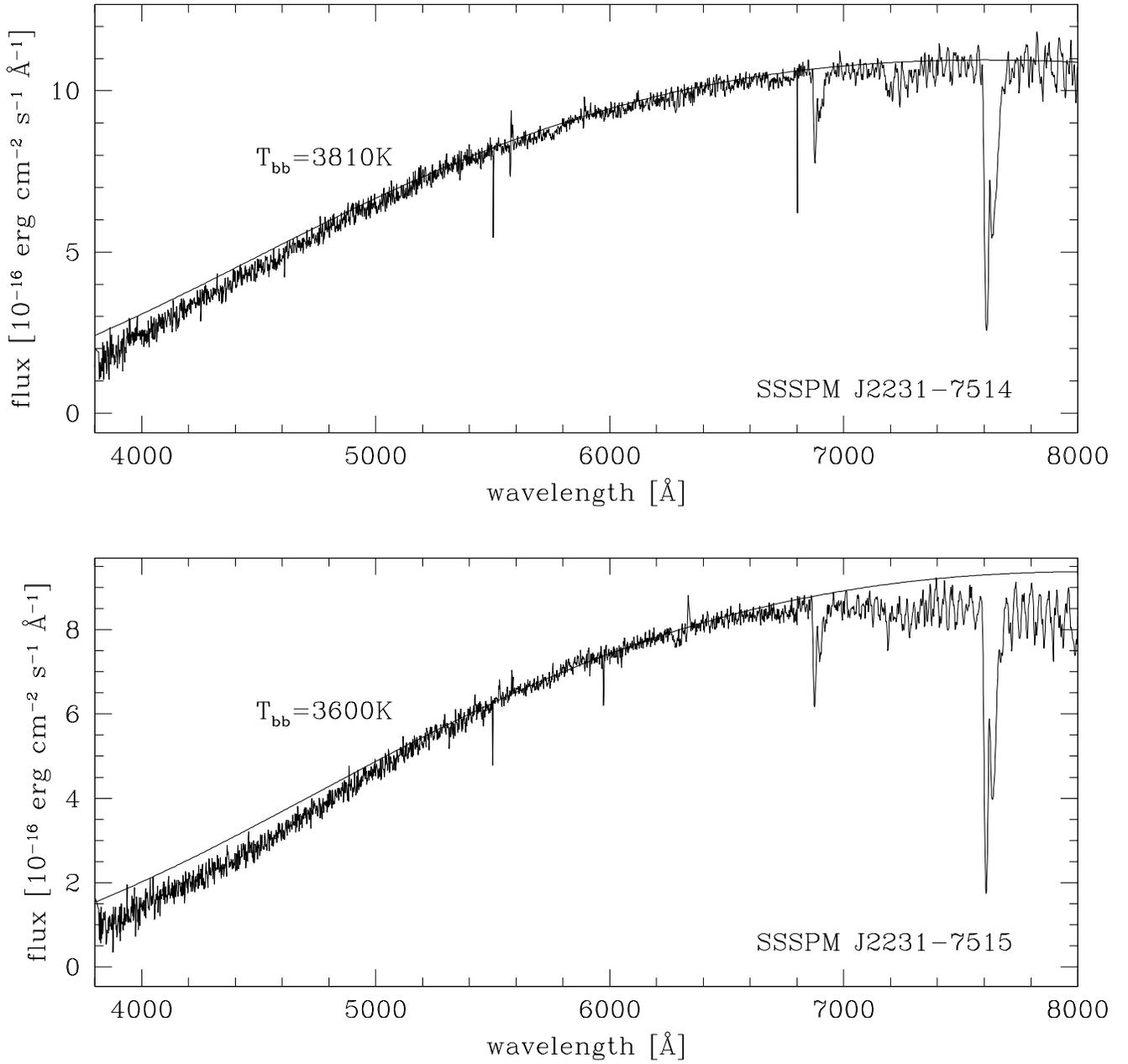}
\caption{Flux calibrated spectra obtained with the Danish 1.5m telescope 
using DFOSC. The atmospheric A- and B-band, which we did not remove, are
the only features seen in the spectra. The red part of the spectra 
($>7000${\AA}) suffers from fringing. There are no spectral features
which could be associated with the objects. We classified both of 
them as extremely cool white dwarfs. Black-body fits to the spectra 
in the range 4300-6800{\AA} (shown by overlaid solid lines) 
yielded temperatures of
3810~K and 3600~K, respectively for the bright and faint 
component.\label{ssspm_spec}}
\end{figure}

The resulting spectra are shown in Figure~\ref{ssspm_spec}. Assuming that our
correction was right, fluxes are accurate to 5-10\%, but in any case,
not worse than 40\% (which {\em can} be a colour dependent effect). As
all our corrections were smooth and the examination of the extracted
raw spectra showed that this not well understood effect is also smooth,
we are quite certain that we neither introduced nor removed fine
structure (e.g. emission or absorption lines) in the data.

Both spectra were fitted with a black-body spectrum in the range 4300-6800{\AA}
(where our data are the most accurate). The fit values are 3810~K for the
bright one and 3600~K for the faint one. If our reduction truly reproduced
the correct continuum shape, the accuracy of these numbers is $\pm$100~K.
As we are not sure that we did not introduce a tilt in the spectra,
we can only say with high degree of confidence that the temperatures are
in the range of 3000-5000~K (clearly cool white dwarfs).

\subsection{Reduction of EFOSC data}

\begin{figure}
\plotone{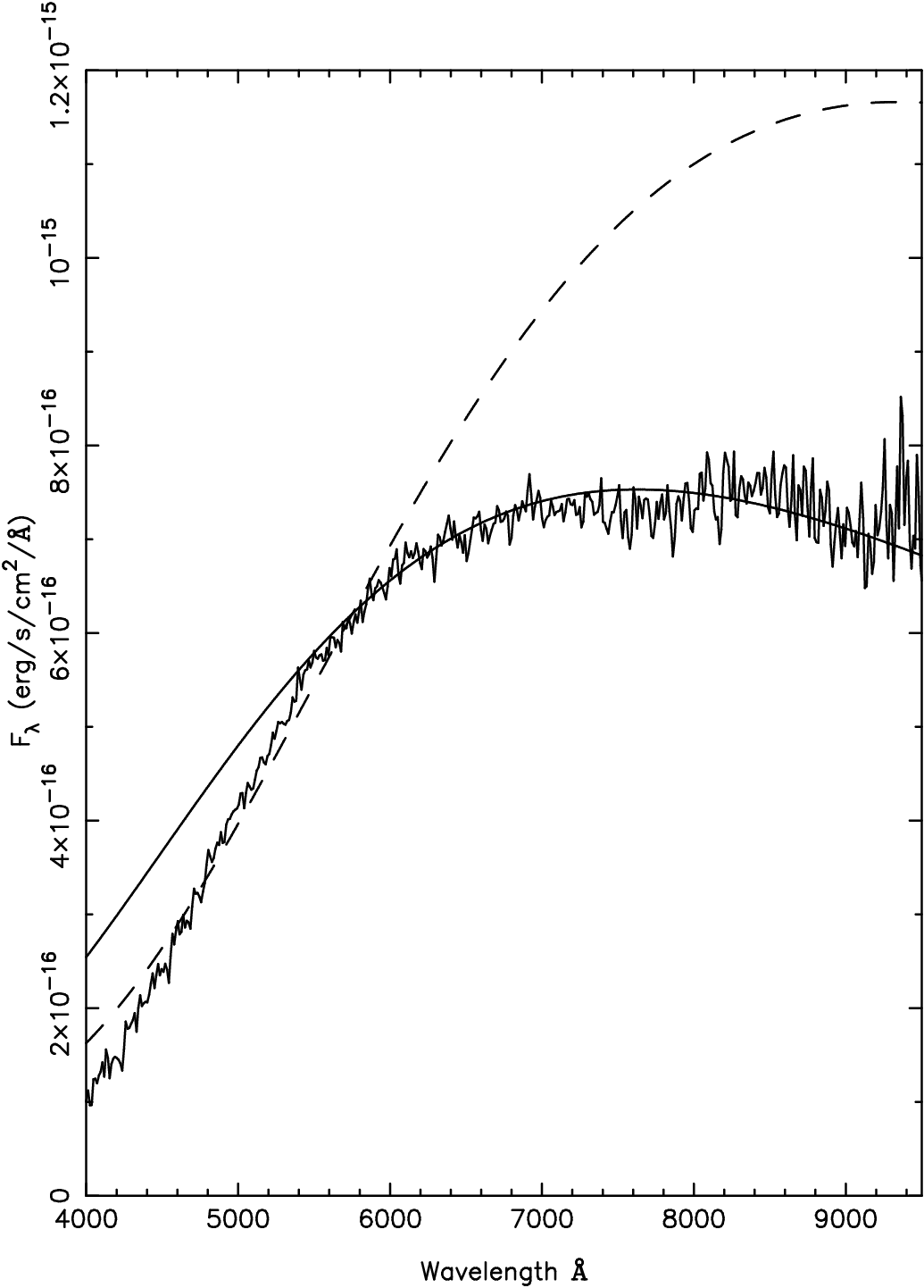}
\caption{Flux calibrated spectrum of the fainter component, SSSPM~J2231-7515,
obtained with EFOSC at the ESO 3.6m telescope. The solid
line shows a black-body fit ($T=3800$~K) to the red side ($\lambda>6000${\AA}), 
which fails to follow the blue side of the spectrum, whereas the dashed line
represents a black-body fit ($T=3100$~K) to the blue end ($\lambda<6000${\AA}),
which gives a poor fit in the red.\label{ssspm_fig3}}
\end{figure}

The three consecutive 300s exposures were combined to 
reject cosmic ray defects, and reduced in a way similar to that detailed above.
The wavelength-calibration (by comparison to He-Ar lines) is accurate to
2.3\AA rms, and flux-calibration was performed by comparison to the
standard star LTT2415. Contrary to the DFOSC observations however, the
EFOSC spectrograph did not give rise to unexpected variations between
exposures in the red part of the spectrum. The resulting spectrum of
SSSPM~J2231-7515 is shown in Figure~\ref{ssspm_fig3}.

We have fit two black-body curves to the spectrum of Figure~\ref{ssspm_fig3}. 
The solid line shows a fit to the red side ($\lambda>6000${\AA}), which has a
temperature of $T=3800$~K, but which fails to follow the blue side of the
spectrum. Fitting instead the blue end ($\lambda<6000${\AA}), gives a cooler
temperature of $T=3100$~K (dashed line), but which gives a poor fit in the
red.  The study of cool white dwarf atmospheres by \citet{ber97} 
employed H- and He atmospheres to fit $BRIJHK$
colors and H-alpha lines. They also often found that the observed $B$-band
flux was too faint for the best fit to $RIJHK$. This would suggest that an
additional source of opacity exists near the B-band, and that the higher
temperature value ($T=3800$~K) is a better estimate for the true
atmospheric temperature.

There is a good agreement of the latter temperature value with that obtained
from the black-body fit to the DFOSC spectrum of SSSPM~J2231-7515
(Figure~\ref{ssspm_spec}). This confirms the corrections applied
in the flux calibration of the DFOSC data and suggests both white dwarfs have 
equivalent black body temperatures $<$4000~K.

\section{Distance estimates}

\subsection{Distance estimate based on magnitude and colour}

With the above temperatures, respectively of 3810~K and 3600~K 
(3100~K to 3800~K), the newly discovered objects
are comparable to the coolest known white dwarf WD0346+246 with 3750~K
\citep{hod00,opp01a} for which a trigonometric parallax of 36$\pm$5~mas
(28$\pm$4~pc)
has been measured \citep{ham99}. If we assume our objects to have the same
physical properties (temperature, mass, chemical composition)
as WD0346+246, we can estimate their distance from a comparison of their
apparent $V$ magnitudes. With $V=19.06$ \citep{opp01a}, WD0346+246 is more than
two magnitudes fainter than our objects (16.60 and 16.87), and consequently
we get distance estimates of 9~pc and 10.2~pc. These distance estimates 
have the
same relative uncertainty as for the comparison object, i.e. $\pm$1.3~pc and
$\pm$1.5~pc, respectively, and rely on the
assumption of identical physical properties, which is unlikely to be the
case.

For an alternative distance estimate we can apply the photometric parallax 
relation given in \citet{opp01b}:
\begin{equation}
M_{B_J}=12.73+2.58(B_J-R).
\end{equation}
With the $B_J$ magnitudes and $B_J-R$ colours from Table~\ref{sssdata}
we get 14.1~pc and 13.1~pc, respectively for the bright and faint component
of our cool white dwarf pair with an uncertainty of about 20\%.
All objects
listed in Table~1 of \citet{opp01b}, with comparable colour ($B_J-R>1.5$)
to our newly discovered objects, are at least two magnitudes fainter in
$B_J$. The fainter component, SSSPM~J2231-7515 is with $B_J-R=+1.75$ and
$R-I=+0.63$ very close to the colours of the two known extremely cool
white dwarfs F351-50 and WD0346+246 as shown in the colour-colour-diagram
of \citet{opp01b}.

\subsection{Distance estimate based on kinematics and separation}

From the angular separation we can make an estimate of the expected 
differential proper motion due to orbital motion of the wide binary if we 
make plausible assumptions for the total mass of the system and the 
orbital characteristics.  In Table~\ref{pmorb} we give an example of the 
results for  masses of both components respectively of 0.4, 0.6 and 0.8 
solar masses in an assumed circular orbit perpendicular to the line of sight.

A conservative estimate of the proper motion errors based on the formal
errors and on the change in proper motion with different numbers of 
distinct epoch positions included in the solution (see Table~\ref{sssdata}),
gives 8~mas/yr.  Since we measure zero differential proper motion within 
these errors, we can conclude that for a circular orbit as above the distance 
must be at least about 15~pc even allowing for some inclination of the orbit. 
Only in the case of an almost edge-on orbital plane with respect to 
the line of sight and both components located such that their orbital
motion manifests itself mainly in line-of-sight radial velocity, would we see
no differential proper motion for smaller ditances.

We have also computed possible heliocentric space velocities
based on assumptions on distance and radial velocity of the cool white
dwarf pair. The distance estimates of about 10~pc to 15~pc, based on 
magnitude and colour and on the angular separation and the not measurable
differential proper motion, already lead to rather large space velocities. 
However, clear halo-like space velocities are only computed assuming 25~pc
to 30~pc distance, which is probably not realistic given the
magnitudes and temperature estimates.


\section{conclusions}

A common proper motion pair with the very large proper motion of about
1.9~arcsec has been discovered from SuperCOSMOS Sky Survey data. Both
objects, SSSPM~J2231-7514 and SSSPM~J2231-7515, were classified from 
follow-up spectroscopy as featureless cool white dwarfs. 
Effective temperatures of both objects were obtained
by fitting the flux-calibrated spectra with black-body models.
If the temperature estimates are correct, the newly discovered white
dwarfs belong to the small number
of presently known extremely cool white dwarfs with $T_{eff}<4000$~K.
With a distance of about 10~pc to 15~pc, obtained from comparison with other
known cool white dwarfs, they would also be the nearest objects of
that temperature class. A conservative temperature estimate of 
$T_{eff}<5000$~K still makes the new objects interesting for follow-up
observations, since they would be only the second known double degenerates in
that temperature class within about 25 pc (after LHS~239 and LHS~240).
Currently, there are only 11 cool white dwarfs with $T_{eff}<5000$~K
and trigonometric parallaxes of less than 25~pc \citep{ber01}. All
presently known or suspected degenerates with $T_{eff}<4000$~K are at
trigonometric or photometric distances of more than 25~pc
\citep{ham99,iba00,har01,opp01b}.

Further investigation of the new objects is obviously needed. A
trigonometric parallax measurement and further improvement of the
(differential) proper motion analysis of the wide binary
will allow a mass estimate. The trigonometric parallax will also
be important for the classification of the new pair of cool white 
dwarfs as Galactic disk or halo components. From the current 
estimates, based on the available data and some assumptions,
the objects might be among the coolest and oldest Galactic disk
white dwarfs, although we can not yet exclude the possibility
that they belong to the dark matter component of the Galactic halo. 
Accurate photometry, including IR photometry is needed to constrain the
amount of absorption due to molecular hydrogen. Our first attempts to
estimate the temperature by black-body fits provided only preliminary
values. As shown by \citet{opp01a}, the physical parameters of cool
white dwarfs can only be obtained from fitting synthetic spectra to the
observed spectral energy distribution. Since the objects are 
rather bright, higher resolution spectroscopy can be carried out
and one can attempt to measure their radial velocity if any lines
can be resolved.

\acknowledgments

This research has made use of the SuperCOSMOS Sky Surveys, i.e. digitized 
data obtained from scans of UKST and ESO
Schmidt plates. We would like to thank the SuperCOSMOS team for producing 
such excellent data. We thank Nigel Hambly for helpful discussion and
for providing us with the $I$ band SSS data.

RDS gratefully acknowledges financial support from the Deutsches Zentrum
f\"ur Luft- und Raumfahrt (DLR) (F\"or\-der\-kenn\-zeichen 50~OI~0001).
GPS acknowledges support under DLR grant 50 OR 9908.
MA gratefully acknowledge support from the
instrument center IJAF in Denmark.


\begin{deluxetable}{ccccccccc}
\tabletypesize{\scriptsize}
\tablecolumns{9}
\tablewidth{0pc}
\tablecaption{SSS positions, photographic photometry and proper motions\label{sssdata}}
\tablehead{
\colhead{Object} & \colhead{RA and DEC} & \colhead{$B_J$} & \colhead{$V$} & 
\colhead{$R$} & \colhead{$R$} & \colhead{$I$} & \colhead{$\mu_{\alpha}\cos{\delta}$} & 
\colhead{$\mu_{\delta}$} \\
\colhead{SSSPM...} & \colhead{J2000, epoch 1996.775}  & \colhead{UKST} & 
\colhead{CCD} & \colhead{ESO}  & \colhead{UKST} & \colhead{UKST} & \multicolumn{2}{c}{mas/yr}} 
\startdata
J2231-7514 & 22 30 39.613  $-$75 13 49.36 & 17.27 & 16.60 & 15.77 & 15.82 & 15.25 & $+404\pm7$ & $-1824\pm9$ \\
J2231-7515 & 22 30 33.141  $-$75 15 18.43 & 17.83 & 16.87 & 16.08 & 16.08 & 15.45 & $+404\pm5$ & $-1829\pm8$ \\
\enddata

\tablecomments{$V$ magnitudes were obtained from the acquisition
images.  The proper motions given in the table were determined from
the SSS positions at four different epochs: 1977.782 ($B_J$), 1984.782
($R$ (ESO)), 1993.733 ($I$) and 1996.775 ($R$ (UKST)).  If we use the recent
positions of our acquisition images (J2000, epoch 2001.470),
22 30 40.136 $-$75 13 58.08 (SSSPM~J2231-7514) and 22 30 33.688 -75 15 27.10
(SSSPM~J2231-7515) in addition to the SSS positions, we obtain a proper
motion of $+408\pm5, -1828\pm7$~mas/yr and $+409\pm5, -1831\pm5$~mas/yr,
respectively.}

\end{deluxetable}



\begin{deluxetable}{cccccccc}
\tablecolumns{8}
\tablewidth{0pc}
\tablecaption{Expected differential proper motion due to orbital motion\label{pmorb}}
\tablehead{
\colhead{distance}  &  \colhead{separation}  & \multicolumn{2}{c}{$2\times0.4M_{\odot}$} & 
\multicolumn{2}{c}{$2\times0.6M_{\odot}$} & \multicolumn{2}{c}{$2\times0.8M_{\odot}$} \\
      &          &  \colhead{Orbital} & \colhead{$\mu_{orb}$} &  \colhead{Orbital} & 
\colhead{$\mu_{orb}$} &  \colhead{Orbital} & \colhead{$\mu_{orb}$} \\
      &          &  \colhead{period} &         &  \colhead{period}  &
         &  \colhead{period}  &         \\
  \colhead{pc}      &    \colhead{AU}        &     \colhead{yr}  &   
\colhead{arcsec/yr} &     \colhead{yr}   &   \colhead{arcsec/yr} &     \colhead{yr}   &   
\colhead{arcsec/yr}}
\startdata
  5\tablenotemark{a} &  465 &  11000  &  0.052 &   9000 & 0.064 &   8000  &  0.074 \\
 10\tablenotemark{a} &  930 &  32000  &  0.018 &  26000 & 0.023 &  22000  &  0.026 \\
 15\tablenotemark{a} & 1395 &  58000  &  0.010 &  48000 & 0.012 &  41000  &  0.014 \\
 20\tablenotemark{a} & 1860 &  90000  &  0.007 &  73000 & 0.008 &  63000  &  0.009 \\
 25\tablenotemark{a} & 2325 & 125000  &  0.005 & 102000 & 0.006 &  89000  &  0.007 \\
 30\tablenotemark{a} & 2790 & 165000  &  0.004 & 135000 & 0.004 & 117000  &  0.005 \\
 10\tablenotemark{b} & 5270 & 428000  &  0.008 & 349000 & 0.009 & 302000  &  0.011 \\
 10\tablenotemark{c} &  930 &  32000  &  0.003 &  26000 & 0.004 &  22000  &  0.005 \\
 10\tablenotemark{d} &  930 &  15000  &  0.012 &  12000 & 0.016 &  10000  &  0.018 \\
\enddata

\tablenotetext{a}{Computations for the first six rows are based on the measured
angular separation of 93~arcsec, the assumed equal masses for both components
and a circular orbit in the plane perpendicular to the line of sight. For the
results shown in the last three rows, additional assumptions were made.}
\tablenotetext{b}{$80^{\circ}$ inclination of the orbit in North-South direction}
\tablenotetext{c}{$80^{\circ}$ inclination of the orbit in East-West direction}
\tablenotetext{d}{No inclination, but eccentricity of 0.6 with stars at apastron.}

\end{deluxetable}



\end{document}